# Micromagnetics of single and double point contact spin torque oscillators

Gino Hrkac[1], Julian Dean[1], Alexander Goncharov[1], Simon Bance[1], Dan Allwood[1], Thomas Schrefl[1], Dieter Suess[2] and Josef Fidler[2]

[1]Department of Engineering materials, University of Sheffield
Sir Robert Hadfieldbuilding, Mapping Street, S13JD Sheffield, UK

[2]Institue of Solid State Physics, Vienna University of Technology
Wiedner Haupt Strasse 8-10, 1040 Vienna, Austria

(Date: April 1, 2008 [ver 1])

Abstract

In this paper we numerically conduct micromagnetic modelling to optimize computational boundaries of magnetic thin-film elements applicable to single and double point contact spin torque nano-oscillators. Different boundary conditions have been introduced to compensate spin waves reflections at boundaries that are based on extended layers, absorbing boundaries, and focal point methods and are compared with a technique based on scattering theory. A surface roughness boundary model is presented which is modelled according to the Rayleigh criterion to minimize specular reflections at computational boundaries. It is shown that the surface roughness model disperses the reflected spin waves and improves the signal to background noise ratio. The model is tested in comparison to conventional approaches such as extended layer systems, variable damping constant and focal point methods for double point contacts. The surface roughness model gives solutions that are stable in time, in qualitative agreement with experiments and capable to reproduce phenomena such as phase locking in double point contacts.

PACS numbers: 75.30.Dj, 75.40.Mg, 75.70.-i, 75.75.+a, 72.25.Pn,

Spin-torque high frequency microwave nanocontact-oscillators of ferromagnetic thin films have attracted interest as monochromatic onboard gigahertz-frequency microwave sources for integrated electronic circuits. It has been theoretically predicted[1-3] and experimentally observed[4-6] that spin-polarized currents in magnetic nanostructure devices may excite steady state microwave magnetization precessions in ferromagnetic thin films. Recent experiments have demonstrated that microwave oscillations in current driven nanocontact devices can be phase-locked in frequency by varying the bias current through one of the contacts[7, 8]. A lot of work has focused on understanding the interaction of spin-polarized current and ferromagnetic nanostructures[9-11] and several models and studies, analytical and numerical[12-15], have been presented to explain experimental results.

For the design of such magneto-electronic nano-oscillators it is crucial to develop reliable simulation techniques. By comparing experiments with numerical simulations one encounters methodological problems that are specific for the problem investigated and the technique used, such as initial conditions, non-arbitrary geometries, limits in the time evolution of the solution (limits on resolution and stability), and the boundary problem.

Especially in micromagnetic simulations of spin torque oscillators the treatment of spin wave propagation and its reflection at the boundaries impose limits on resolution and stability[13]. Different solutions were suggested such as shapes with focal points[16] or absorbing boundaries[17].

In this paper we propose a novel technique that uses wave scattering at rough surfaces. Based on scattering theory we apply a surface roughness at the boundary, which results in a spatial change of the reflection coefficient, the amplitude of the scattered waves is reduced compared to that at a smooth surface and the reflections are minimized. A detailed micromagnetic study that simultaneously solves the quasi-static Maxwell equations and the Landau-Lifshitz-Gilbert Slonczewski (LLGS) equation for a single and double-point contact regime with different surface configurations is performed. For the single point contact spin valve system we investigate three different boundary models: the absorbing boundary model (based on a gradually increase of the damping constant), a smooth boundary model with a transition area where the exchange is reduced and the spins are weaker coupled, and a surface roughness boundary model which is based on the Rayleigh criterion. For the double point contact we investigate two configurations: an elliptical model with focal points that uses a smooth boundary and a circular model with a surface roughness boundary model. Both models are evaluated based on their validity to reproduce experimental results published by Kaka et al[8].

## 1. THEORY

The solution of an electromagnetic problem including nonlinear magnetization dynamics consists of solving the static Maxwell equations and the modified Landau-Lifshitz-Gilbert Slonczewski equation with appropriate initial and boundary conditions. We use a micromagnetic approach starting from the total magnetic Gibbs free energy $E_{tot}$

$$E_{tot} = \int_V \left( \frac{A}{M_s^2} (\nabla \mathbf{M})^2 + E_{anis} - \mu_0 \mathbf{M} \left( \mathbf{H}_a + \frac{\mathbf{H}_d}{2} \right) \right) dV \qquad (1)$$

which is discretized using a finite element method that is solved using a hybrid finite element/boundary element method[18, 19]. The first term in equation 1 is the exchange energy followed by the anisotropy energy density, the applied field $\mathbf{H}_a$ and the magnetostatic energy with $\mathbf{H}_d$ being the demagnetizing field, with $\mu_0$ being the permeability of free space. The nonlinear magnetization dynamics is described by a modified Landau-Lifshitz-Gilbert equations, which is written in the following dimensionless form

$$(1+\alpha^2)\frac{d\mathbf{m}}{d\tau} = -\mathbf{m}\times\mathbf{h}_{eff} - \alpha\mathbf{m}\times(\mathbf{m}\times\mathbf{h}_{eff}) - \mathbf{N} \qquad (2)$$

with the spin transfer torque **N** describing the effects caused by the spin polarized current. The magnetization vector $\mathbf{m} = \mathbf{M}/M_s$ is the magnetization **M** normalized by the saturation magnetization $M_s$, and is assumed to be spatially nonuniform, α is the Gilbert damping constant, τ the time measured in units of $(|\gamma| M_s)^{-1}$ (γ is the gyromagnetic ratio) and $\mathbf{h}_{eff} = \mathbf{H}_{eff}/M_s$ is the normalized effective field. The effective field is the negative variational derivative of the total magnetic Gibbs free energy density. We perform zero temperature simulations and do not include a random thermal field in the effective field. Experiment and theory show[22] that the full width at half maximum (FWHM) increases with increasing temperature. Therefore the FWHM calculations in this paper are performed to validate the quality of the different model approaches. The FWHM is calculated from a Lorentzian fit to the Fourier transform of the total free layer magnetization in a time window of 0 to 20 ns. The spin transfer torque **N** in equation 2 is given in the asymmetric form [20] and can be written as

$$\mathbf{N} = \frac{\hbar J_e}{2ed}\frac{1}{\mu_0 M_s^2}g_T\mathbf{m}\times(\mathbf{m}\times\mathbf{p}) = \frac{1}{2}\frac{J_e}{J_p}g_T\mathbf{m}\times(\mathbf{m}\times\mathbf{m}_p) \qquad (3)$$

Here $J_e$ is the electron current density, $J_p = (\mu_0 e d M_s^2 / \hbar)$ the characteristic current density of the system (ℏ is the Planck constant, $e$ the electron charge, $d$ the free layer thickness) and $\mathbf{m}_p$ is the unit vector in the direction of the magnetization in the reference layer. For a NiFe free layer with a thickness of 5 nm and a saturation magnetization of $7.16\times10^5$ A m$^{-1}$ one obtains a characteristic current density $J_p$ of 4.9

$\times 10^{12}$ A m$^{-2}$. The scalar function g$_T$ describes the angular dependency of the spin torque term and is commonly defined as follows[21]

$$g_T(\theta) = \left( \frac{8P^{3/2}}{3(1+P)^3 - 16P^{3/2} + (1+P)^3 \cos(\theta)} \right) \quad (4)$$

Where P is the spin polarization factor and cos(θ)=**m·m$_P$** is the projection of the magnetization on the free layer **m** and the fixed layer **m$_P$**. The spin current polarization is expressed by the constant P with values ranging from $0 \leq P \leq 1$. In the following simulation we assumed a value of P = 0.2.

In the following we consider spin torque oscillators in the point contact geometry. The extended GMR spin valve stack consists of a 5 nm NiFe layer, a 5 Cu layer and a 20 nm CoFe layer and is contacted with two Cu point contacts. The whole system is current driven, meaning that the difference in the magnetization between the NiFe and the CoFe layer is seen as a voltage output. Once a spin polarized current is injected into the system using the theory described above it exerts a torque on the localised magnetization and leads to a uniform oscillation below the contacted area. This uniform oscillation creates spin waves that propagate through the thin film to the boundaries of the system. In our micromagnetic model we have to limit the lateral size of the system compared to experiments, which gives rise to artificial interference effects, such as reflections of spin waves at the computational boundary. For this propagation problem we use a geometrical solution coming from optics that is based on diffuse reflection at rough surfaces. Diffuse reflection can be observed when lightwaves are reflected back from a polished surface, like a mirror. On a perfectly smooth surface lightwaves are speculary reflected and give an image of the source. By introducing an increased rugosity at the surface the speculary reflected energy is reduced and is scattered in other directions. The determinant parameter is the ratio of the wavelength of the incident spin wave and the lateral dimension of the geometrical fluctuation at the boundary which is described by the Rayleigh criterion[23].

The Rayleigh criterion distinguishes between rough and smooth surfaces. A surface is considered smooth when the two optical paths of two incident waves reflected on a surface at two points at a distance d are equal and giving coherent waves in the specular direction. By introducing an irregularity between the two points of height h the optical paths are no longer equal. The path difference is

$$\Delta H = 2h\cos(\theta) \tag{5}$$

where θ is the incident angle measured from the surface normal. This path difference results in a phase difference

$$\Delta\Omega = \frac{2\pi\Delta H}{\lambda} = \frac{4\pi h\cos(\theta)}{\lambda} \tag{6}$$

with λ being the wavelength of the incident wave. If the irregularity height h is small the surface is considered smooth and considered rough once h reaches a certain value. The Rayleigh criteria use a threshold value of π/2 for the phase difference. By using this criteria in equation 6 a surface is considered as rough if

$$\frac{4\pi h\cos(\theta)}{\lambda} \geq \frac{\pi}{2} \quad \text{or} \quad h \geq \frac{\lambda}{8\cos(\theta)} \tag{7}$$

## 2. MODEL

The study was carried out on a trilayer spin valve structure consisting of $Co_{90}Fe_{10}$ (20nm)/Cu (5nm)/$Ni_{80}Fe_{20}$ (5nm). The system is either contacted with a single Cu contact (at the centre of the thin film) or two Cu contacts (each 40 nm in diameter) for the double point contact regime in which case the centre to centre distance is 500 nm located along the z axis. The $Co_{90}Fe_{10}$ layer is considered as the fixed layer meaning that the magnetization is pinned in terms of the spin torque driven magnetization processes due to a larger thickness, *d*, and higher saturation magnetization $M_s$ (1.43×10$^6$ A m$^{-1}$) and exchange constant A (2.5×10$^{-11}$ Jm$^{-1}$) compared to the Ni$_{80}$Fe$_{20}$ layer ($M_s$ = 7.16×10$^5$ A m$^{-1}$, A = 1.3×10$^{-11}$ Jm$^{-1}$) here refereed as the free layer.

In the following simulations we applied an external field of 740 mT at 15° out of plane. In the double point contact regime the field projection on the film coincides with the connecting line of the two contacts.

The overall size of the simulation area for the single point contact model was 400 nm in diameter with an additional extended layer with a width of 400 nm (diameter of the simulation area = 1200 nm) for the smooth and the absorbing boundary model. The total diameter for the surface roughness model was 400 nm, see Fig. 1. For the double point contact model the simulation area was 500 nm in radius with an additional extended layer with a width of 500 nm (diameter of the simulation area = 2000 nm) for the focal point boundary model. For the surface roughness model the diameter for the simulation area was 1000 nm. The discretization size is 5 nm which is below the exchange length of 5.5 nm. The system is current driven, meaning that the changes in the relative orientation of the magnetization in the free and fixed layer are seen as voltage changes due to the GMR effect.

## 3. NUMERICAL SIMULATIONS AND RESULTS

In order to identify the problems that arise by using a micromagnetic framework with no special boundary conditions at the computational boundary to describe experimental results, we start the analysis with a double point contact spin valve system as described above with a smooth geometrical boundary. The radius of the system is 850 nm with a discretization size of 5 nm and an outer ring layer with a width of 250 nm which has an abrupt increased discretization size from 5 to 20 nm. A current density of $6.37 \times 10^{12}$ Am$^{-2}$ is applied through one of the contacts and the simulation time was 20 ns.

By applying a spin polarized current through the point contact, spin waves with a wavelength of 55 nm are emitted and propagate homogenously through the thin film toward the boundary. At the boundary the spin waves are partially dampened and partially reflected back into the system. This can be explained by the gradually increase of the mesh size that leads to a decrease of the effective exchange constant[24] of the magnetization vectors once the mesh size is bigger than the exchange length. In addition the abrupt change in mesh size is seen by the spin wave as an artificial boundary. Due to this transition area two effects can be observed in the Fourier space: first an increase in the background noise, see Fig. 2, and secondly additional peaks due to mirror current sources, see Fig. 2.

From these results one sees that this boundary approach is not sufficient and accurate enough to derive comparable results with experiments.

### A. Single point contact

For the single point contacted spin valve system we investigate the influence of three different surface configurations (see Fig. 1) on the spin wave propagation and the stability of the oscillation behaviour of the spin valve structure.

The first configuration consists of a smooth surface with an additional absorbing area surrounding the simulation area, see Fig. 1. The extended layer consists of three subsections were the damping constant is gradually increased from 0.01 to 0.1 and 0.5. The gradient increase of the damping constant is an artificial absorbing boundary condition used to reduce the spin wave reflections at the computational boundary in such a way that once the spin waves enter the area with increasing damping the magnetization orients itself faster along the effective field direction and therefore dampening the spin wave propagation. The initial magnetization configuration is chosen in such a way to reach a stable oscillation path in finite time. Therefore the magnetization of the free layer and of the reference layer was set perpendicular to each other. Spin waves are injected into the system by applying a current density of $6.37 \times 10^{12}$ Am$^{-2}$ through the point contact. After an initial chaotic state, the magnetization reaches an oscillation cycle that remains stable for 8ns. Starting at 8 ns the steady oscillation below the point contact becomes unstable due to reflected spin waves, see Fig. 3. Looking at the spatial configuration of the magnetization (yz-plane) it can be observed that at the interface where the damping is increased "pinning sites" for spin waves are formed. Due to the increased damping, once the magnetization reaches the equilibrium state in the direction of the effective field it remains constant. This acts like an artificial boundary and therefore increases the reflections. The reflected spin waves build up over time and lead to a destructive interference that destroys the uniform oscillation behaviour below the point contact, see Fig. 3.

The second configuration consists of the freelayer and an extended layer attached to it (see Fig. 2(b)) with a transition area where the meshing is gradually increased from 5 to 20 nm. The width of the extended layer is 400 nm and the decrease in meshing quality reduces the exchange between the individual magnetic spins as the exchange length for the material parameters used in this model is 5.5 nm. The decoupling of the magnetic spins dampens the spin waves and the reflections are reduced. The reduction of the spin waves reflections lead to a uniform and stable oscillation over 18 ns, see Fig. 3(a). However the spin wave reflections build up over time and at 19 ns the reflected spin waves start to interfere with the uniform oscillation, seen in the amplitude fluctuations of the magnetization in y, see Fig. 3(b), and in the spatial configuration of the magnetization (Fig. 3(d)). The Fourier analysis of the magnetization give a dominant frequency peak at 25.25 GHz and a full width at half maximum (FWHM) of 63 MHz. Furthermore the reflected spin waves contribute to a higher background noise and a background tail just before the dominant peak, see Fig. 4. This background tail encompasses the Fourier signals from the reflected spin waves. The oscillation path of the average magnetization of the contacted area is shown in Fig. 5 as a three dimensional trajectory of the magnetization over 20 ns. It can be seen that the oscillations path is fluctuating in time which is shown by the projections of the magnetization onto their respective planes. This can be explained by the fact that the reflected spin waves interfere with the uniform magnetization and lead to a distorted oscillation path which is also shown by the high FWHM of 63 MHz, see Fig. 4.

For the third configuration we use a boundary method that is based on scattering theory described in the theory section of this paper. We use the Rayleigh criterion to design a freelayer with a surface roughness to minimize specular reflections, see Fig. 1. The average radial irregularity height is 12.5 nm and the pattern is based on a distorted toothed wheel, see Fig. 6. The roughness is designed in such a way that the incident angle of the spin wave to the surface normal varies from $\theta_1$ 3° to $\theta_2$ is 52°, see Fig. 6. The spin wave wavelength is 55 nm and the Rayleigh criterion demands for an angle of 3° an irregularity height of at least 6.6 nm and for 52 ° a minimum height of 11.16 nm which is satisfied by the irregularity height of 12.5 nm. Meaning that the surface can be considered rough, and the specular reflections are reduced to a minimum. The time evolution of the averaged magnetization in y shows a stable oscillation over the whole simulation time with no variations in the amplitude, see

Fig. 3(a) and 3b. The snapshot of the spatial configuration shows spin waves originating from the point contact with only small distortion at the edges, see Fig. 3(e). The Fourier transform of the magnetization in y gives a dominant frequency peak at 24.39 GHz with no background tail compared to the smooth boundary model which is in good agreement with experiment[5]. The FWHM of this peak is 33 MHz which is 2 times smaller compared to the other model. The oscillation path of the average magnetization of the contacted area over 20 ns is shown in Fig. 5 and it can be seen that the path is highly uniform and that due to the diffuse reflections the distortions are reduced to a minimum.

## B. Double point contact

For the double point contact spin valve system we investigate the influence of two different surface configurations on the spin wave propagation and the stability of the oscillation behaviour of the spin valve structure with the aim to simulate phase locking between two point contacts, as observed by Kiselev[4].

The first spin valve system has an elliptical geometry with a smooth focal point boundary model and an extended layer surrounding it. The major and minor axis of the ellipse are 2a=1000 nm and 2b=700 nm, respectively and the external field of 740 mT and 15 degree out of plane is applied along the major axis which coincides with the connecting line of the two Cu point contacts, each 40 nm in diameter. The width of the extended layer surrounding the ellipse is 500 nm and the mesh size is gradually increased from 5 nm to 20 nm. The reason for an elliptical geometry is to have the specular reflexions of the spin waves in two focal points which are located 357 nm on both sides from the centre of the ellipsoid and do not coincide with the point contacts which are 250 nm from the centre of the ellipsoid to reduce the destructive interference below the point contacts. We choose the same initial magnetization configuration as described for the single point contact case and apply a current density of 7.16x10$^{12}$ Am$^{-2}$ through both point contacts. Both contacts start to emit spin waves with a wavelength of 55 nm. The waves propagate through the thin film and once they hit the computational boundary, where the mesh size is gradually increased from 5 to 20 nm, the waves get partly transmitted and reflected. The transmitted waves in the extended layer are dampened as the mesh size becomes larger than the exchange length which weakens the exchange between the magnetization vectors. The reflected spin waves are reflected back into two focal points. As the focal points do not coincide with the position of the source of the spin waves the reflected spin waves don't give rise to a destructive interference below the point contacts.

A stable oscillation can be observed over a time scale of 10 ns, see Fig. 7. But as the spin waves propagate through the spin valve system they still contribute to the nonlinear magnetization behaviour which can be seen in the snapshot of the spatial configuration of the magnetization of the focal point model at 14 ns, see Fig. 7. This and the fact that the external field is applied along the connecting line of the point contacts in the positive z-direction which imposes an asymmetry on the propagation of the spin waves, favouring the ones travelling in the positive z-direction and damping the ones travelling in the opposite direction, leads to a breakdown of the uniform magnetization at point contact **a** and a stable oscillation below point contact **b**, see spatial configuration of magnetization at 20 ns, Fig. 7. The Fourier analysis of the magnetization gives a dominant frequency peak at 24.84 GHz and at 25.07 GHz, with a full width at half maximum of 79 MHz and 98 MHz, respectively, see Fig. 8. The reflected spin waves contribute to a higher background noise and to a front and backward tailing of the dominant peaks that encompass the Fourier signals from the reflected spin waves.

The second system has a circular geometry with a radius of 600 nm and uses a surface roughness boundary model. The two point contacts (40 nm in diameter) are located symmetrically (250 nm) around the centre of the circular spin valve system and along the z-axis. There is no extended layer and the mesh size for this system is 5 nm. The same initial conditions for the magnetization and for the external field are used as for the first double point contact model. A current density of $7.16 \times 10^{12}$ Am$^{-2}$ is applied through both point contacts. The contacts start to emit spin waves with a wavelength of 55 nm. The waves propagate to the computational boundary and get scattered according to the Rayleigh criterion (same conditions as for the single point contact system). A stable oscillation behaviour can be observed over the entire simulation time, see spatial configuration of magnetization at 10, 14 and 20 ns, Fig. 7. The Fourier analysis of the magnetization gives a dominant frequency peak at 24.49 GHz and at 25.10 GHz, with a full width at half maximum of 46 MHz and 56 MHz, respectively, see Fig. 8. The scattering of the spin waves result only in a small increase of the background noise. This can be seen by comparing the power output of the elliptical (low signal to background noise ratio) and the surface roughness model (high signal to background noise ratio), see Fig. 8. But around 28 GHz also for the surface roughness an increase in the background noise can be observed as in the elliptical model.

Although the waves are scattered, the energy of the spin waves is still confined in the system and is shown by an increase of the heterogeneous magnetization configuration close to the computational boundary, see Fig. 7, which leads to a broadening of the FWHM, see Fig. 8.

The surface roughness model gives a stable solution in time and the results are in qualitative agreement with experiments[5, 6]. Once the model was validated, the current through one of the point contacts was increased to observe a phase locking behaviour as seen by Kiselev and Kaka[5, 8]. In our system the current density was gradually increased in contact **b** and the current density in contact **a** was fixed at $7.16 \times 10^{12}$ Am$^{-2}$. The simulations show a red shift in frequency by increasing the current density and once the current density in contact **b** has reached a value of $6.37 \times 10^{13}$ Am$^{-2}$ a locking of the frequencies over a current range of 2 mA can be observed. Figure 9 shows the Fourier analysis of the magnetization in y for two current density configurations, one were the current density through both contacts is $7.16 \times 10^{12}$ Am$^{-2}$ and the other where the current density through contact **a** is $7.16 \times 10^{12}$ Am$^{-2}$ and $6.37 \times 10^{13}$ Am$^{-2}$ in contact **b**. The Fourier analysis in Fig. 9 shows for the first configuration two dominant peaks, one at 24.49 GHz and the other at 25.10 GHz, with a full width at half maximum of 46 MHz and 56 MHz, respectively. For the second configuration only one peak can be observed at 24.50 GHz with a FWHM of 45 MHz. The power of the phase locked peak is twice as much compared to the two separated peaks. This and the fact that the frequency doesn't change over a current increase in the range of 2 mA is strong indication for a phase locking behaviour as described by Kaka et al[8]. More details on the micromagnetics of phase locking in double point contacts will be published elsewhere[25].

## 4. CONCLUSIONS

In summary we presented a novel technique to suppress spin wave reflections at computation boundaries within the framework of micromagnetics. The technique we present is based on scattering theory and the Rayleigh criterion is used to design boundaries that minimize reflections of incident spin waves. The new model was tested in comparison to conventional approaches such as extended layer systems, variable damping constant and focal point methods for double point contacts. The surface roughness model gives solutions that are stable in time and can produce phenomena such as phase locking in double point contact regimes. The results presented in this paper are in qualitative agreement with experiments[4-8].

## ACKNOWLEDGMENTS

This work was supported by the European Communities programs IST STREP, under Contract No. IST-016939 TUNAMOS.

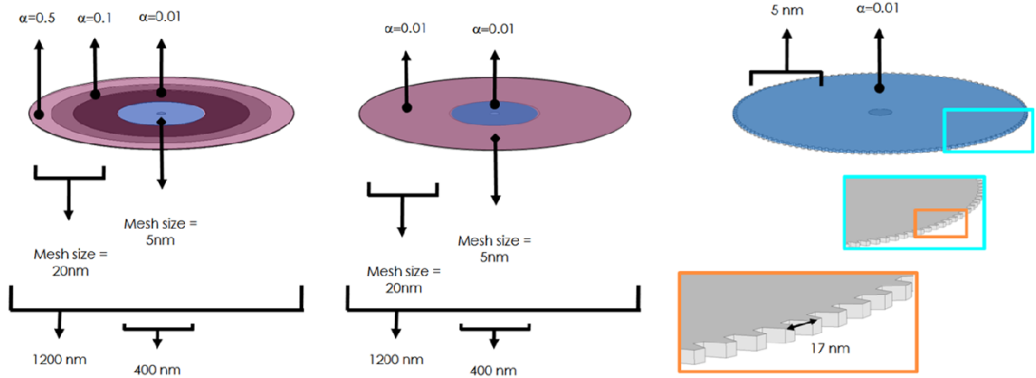

FIG. 1. (Color online) Geometry of a single point contact spin valve model with lateral dimensions, mesh size and damping values. Right: absorbing boundary model. Middle: smooth boundary model with transition area. Left: surface roughness model with an average irregularity height of 12.5 nm.

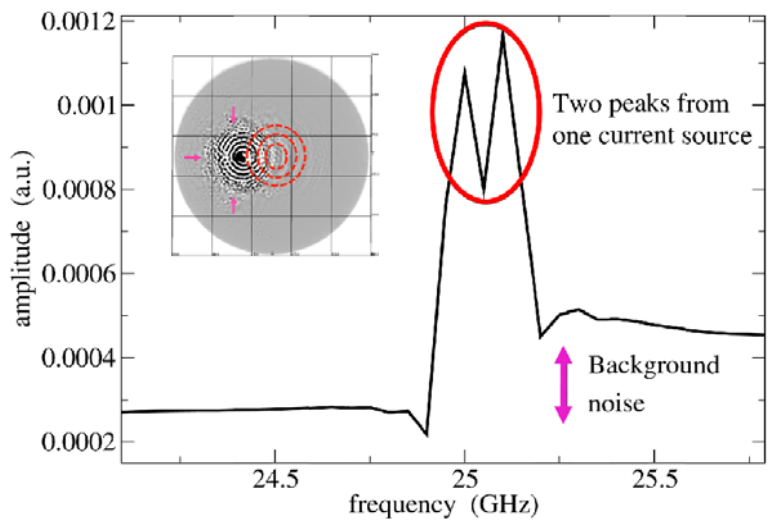

FIG. 2. (Color online) Fourier spectrum of the magnetization in y. The mirror source results in an extra peak and the reflections at the edges (arrows) lead to an increase in the background noise. Inset: spatial configuration of magnetization (yz-plane) of a double point contact with a gradually increased mesh size at the boundary between inner and outer layer (arrows). The dotted line circles show the propagating spin waves from the mirror source.

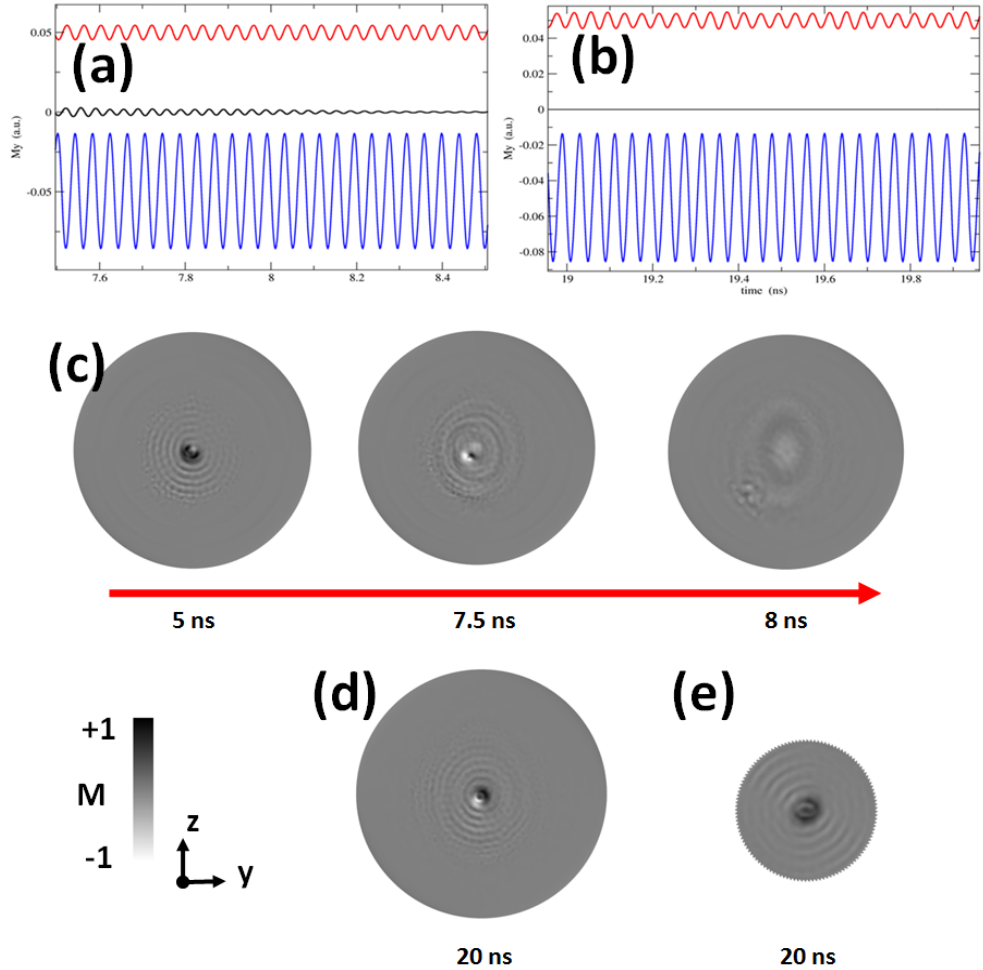

FIG. 3. (Color online) (a) and (b) Time evolution of the normalized and contact averaged magnetization in y. The signal of the smooth boundary model (red) is shifted from the zero centre line to + 0.05 for better comparison. The signal from the absorbing model is around the zero centre line (black) and the signal from the surface roughness boundary model (blue) is shifted by -0.05. (c) Shows snapshots of the spatial configuration of the magnetization of the absorbing boundary model for three time steps 5, 7.7, and 8ns. Starting with nice distinguishable spin waves (5 ns) originating from the point contact. At 7.5 ns the spin waves become more non uniform due to reflections and at 8 ns the spin waves disappear due to destructive interference. At the left lower quarter of the 8 ns snapshot a mirror source can be seen. (d) Spatial configuration of the magnetization for the smooth boundary model at 20ns. At the edges non uniformities in the magnetization can be seen. (e) Spatial configuration of the surface roughness model at 20 ns with clear distinguishable spin waves originating from the point contact.

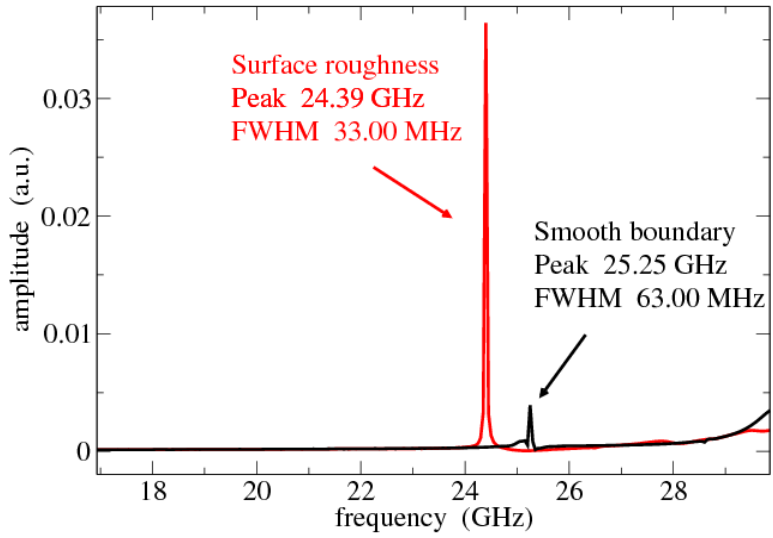

FIG. 4. (Color online) Fourier analysis of the magnetization in y, amplitude as function of frequency for a current density of $6.37 \times 10^{12}$ Am$^{-2}$ and an external field of 740 mT applied 15 degree out of plane: (Left) dominant peak of the surface model at 24.39 GHz and a FWHM of 33.00 MHz; (Right) dominant peak of the smooth boundary model with a forward tail at 25.25 GHz and a FWHM of 63 MHz

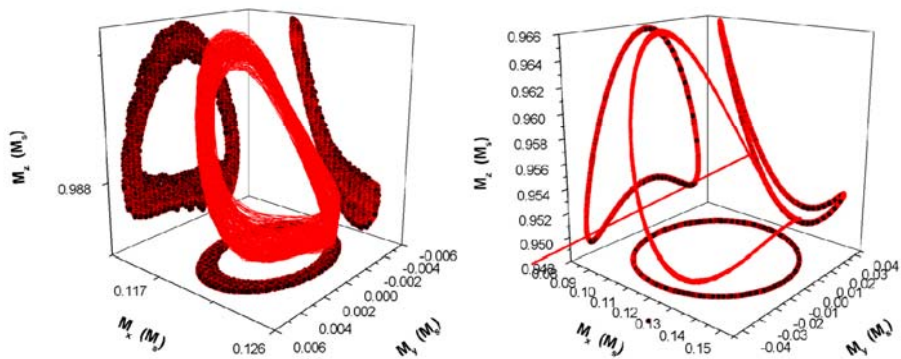

FIG. 5. (Color online) Left: Oscillation path of the smooth boundary model as a 3D trajectory of the magnetization averaged over the contact area over 20 ns. Right: Oscillation path of the surface roughness model.

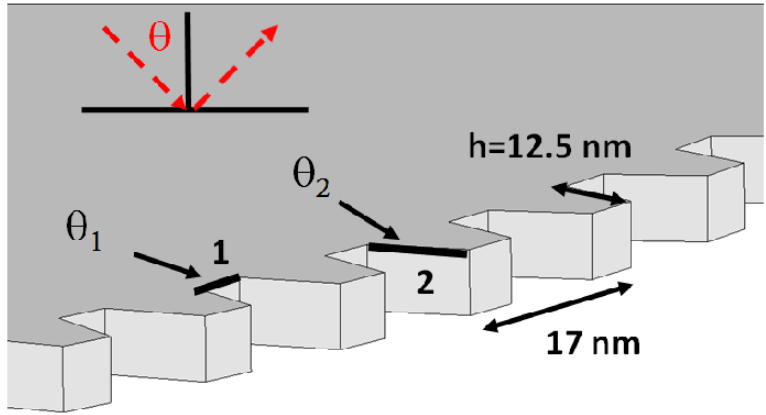

FIG. 6. (Color online) Snapshot of the surface roughness boundary model with an irregularity height h of 12.5 nm. $\theta_1$ is the incident angle of the spin wave in respect to the surface normal of surface 1 and $\theta_2$ is the incident angle in respect to the surface normal of surface 2.

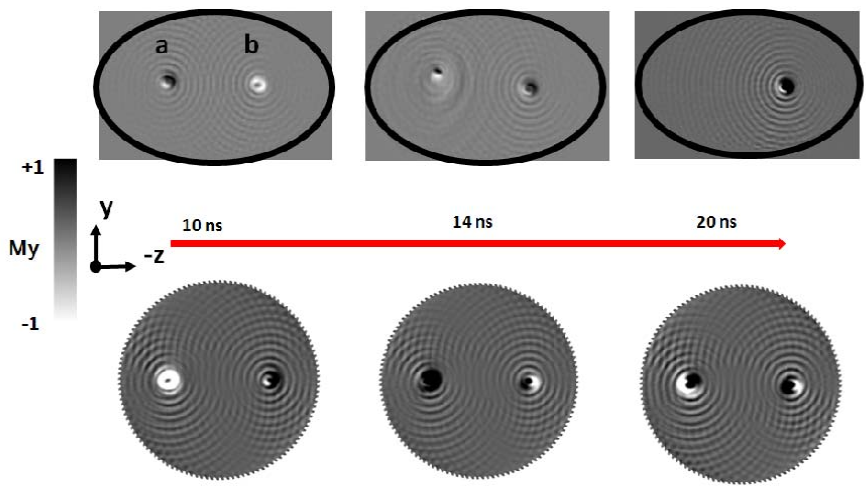

FIG. 7. (Color online) Snapshots of the spatial configuration of the magnetization of the elliptical focal point boundary and the surface roughness boundary model for three time steps 10, 14, and 20ns, with a current density of $7.16 \times 10^{12}$ Am$^{-2}$ through both contacts and an external field of 740 mT applied 15 degree out of plane in the positive z-direction: (Top: elliptical focal point boundary model) Starting with nice distinguishable spin waves (10 ns) originating from both point contacts (**a** and **b**). At 14 ns the magnetization behaviour becomes non uniform. At 20 ns the oscillations disappear at point contact **a** but stable oscillations can be observed for point contact **b**. (Bottom: surface roughness boundary model) Spatial configuration of magnetization shows for the whole simulation time clear distinguishable spin waves originating from both point contacts.

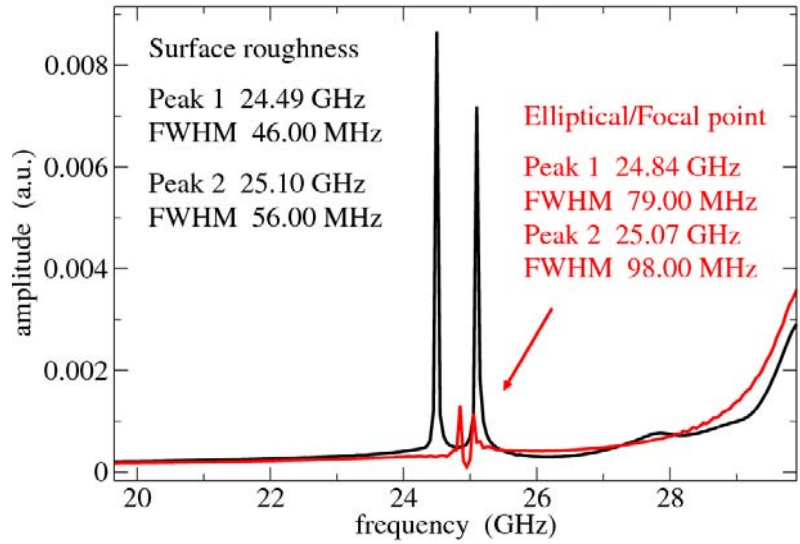

FIG. 8. (Color online) Fourier analysis of the magnetization in y, amplitude as function of frequency for a current density of $7.16 \times 10^{12}$ Am$^{-2}$ through both contacts and an external field of 740 mT applied 15 degree out of plane: two dominant peaks of the surface roughness model at 24.49 GHz and 25.10 GHz and a FWHM of 46.00 MHz and 56.00 MHz; two dominant peaks of the smooth boundary model with a forward and backward tail at 24.84 GHz and 25.07 GHz and a FWHM of 79.00 MHz and 98.00 MHz are observed.

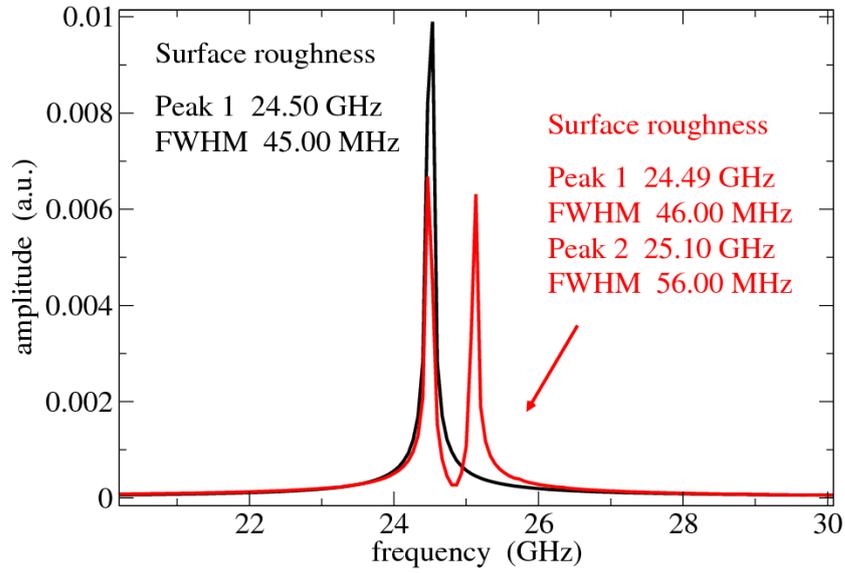

FIG. 9. (Color online) Fourier analysis of the magnetization in y, amplitude as function of frequency for a current density of $7.16 \times 10^{12}$ Am$^{-2}$ through both contacts (red) and an external field of 740 mT applied 15 degree out of plane and for a current density of $7.16 \times 10^{12}$ Am$^{-2}$ though contact a and $6.37 \times 10^{13}$ Am$^{-2}$ through contact b (black): two dominant peaks one at 24.49 GHz and the other at 25.10 GHz with a FWHM of 46.00 MHz and 56.00 MHz; one dominant peak 24.50 GHz and a FWHM of 45.00 MHz are observed.